\documentclass[prl,aps,twocolumn,nofootinbib,10pt, showpacs]{revtex4-1}\usepackage{amsmath}
\usepackage{graphicx}
\usepackage{mhchem} %[version=4]
\usepackage{braket}         

\begin{document}

\title{Topological spin Hall effect in antiferromagnetic skyrmions}

\author{Patrick M. Buhl}
\email{p.buhl@fz-juelich.de}
\author{Frank Freimuth}
\author{Stefan Bl\"ugel}
\author{Yuriy Mokrousov}
\affiliation{Peter Gr\"unberg Institut and Institute for Advanced Simulation, Forschungszentrum J\"ulich and JARA, 52425 J\"ulich, Germany}


\begin{abstract}
  The topological Hall effect (THE), as one of the primary manifestations of non-trivial topology of chiral skyrmions, is traditionally used to detect the emergence of skyrmion lattices with locally ferromagnetic order. 
  In this work we demonstrate that the appearance of non-trivial two-dimensional chiral textures with locally {\it anti}-ferromagnetic order can be detected through the spin version of the THE $-$ the topological spin Hall effect (TSHE).  
  Utilizing the semiclassical formalism, here used to combine chiral antiferromagnetic textures with a density functional theory description of the collinear, degenerate electronic structure,   we follow the real-space real-time evolution of electronic SU(2) wavepackets in an external electric field to demonstrate the emergence of sizeable transverse pure spin current in synthetic antiferromagnets of the Fe/Cu/Fe trilayer type.
  We further unravel the extreme sensitivity of the TSHE to the details of the electronic structure, suggesting that the magnitude and sign of the TSHE in transition-metal synthetic antiferromagnets can be engineered by tuning such parameters as the thickness or band filling. 
  Besides being an important step in our understanding of the topological properties of ever more complex skyrmionic systems, our results bear great potential in stimulating the discovery of antiferromagnetic skyrmions.
\end{abstract}

\date{\today}
\maketitle

{\bf Introduction}. The discovery of chiral skyrmions in metallic compounds such as MnSi \cite{Pfleiderer2009} has triggered an enormous theoretical and practical interest in these topological chiral textures which exhibit such fascinating properties as topological transport, fast dynamics and topological robustness.  
While diverse applications of skyrmions in spintronics are currently discussed, many of them rely on an idea of skyrmion propagation in a device realizing either information transfer or racetrack memory. Conventional, locally ferromagnetic skyrmions, found~e.g.~in FeGe or at the interfaces between Co and Pt \cite{Yu2011,Moreau-Luchaire2016,Soumyanarayanan2016,Woo2016,Dupe2016}, suffer from the skyrmion Hall effect which intrinsically roots in a Magnus force pushing the skyrmions towards the edges of a device and thus greatly limiting their range of possible applications \cite{Purnama2015}. 
Recently, based on micromagnetic simulations, it was suggested that this problem could be circumvented in chiral skyrmions which exhibit a local {\it anti}-ferromagnetic coupling between the spins \cite{Zhang2016,Zhang2016a}. It was argued that not only in such antiferromagnetic (AFM) skyrmions the Magnus forces on each of the staggered magnetic sublattices cancel each other, but also that the velocity of AFM skyrmions along an applied electric field would be significantly faster than that of their ferromagnetic twins \cite{Zhang2016,Zhang2016a,Velkov2016}. However, despite bright
prospects of antiferromagnetic textures and AFM materials in general~\cite{Wadley2016,Jungwirth2016}, the experimental observation of AFM skyrmions has not been reported so far. This can be attributed
to the fact that the AFM structures are much less sensitive to external magnetic fields, while the detection
of chiral antiferromagnetic textures is very challenging owing to the staggered nature of the local spin
distribution \cite{Wiesendanger2009}.

One of the key manifestations of non-trivial topology of locally ferromagnetic skyrmions is the so-called  topological Hall effect (THE) \cite{Bruno2004,Nagaosa2013,Denisov2016}.
Unlike the ordinary and anomalous Hall effects, the THE within the adiabatic picture originates in the skyrmion's spin structure giving rise to an ``emergent" magnetic field which exerts a spin-dependent Lorentz force on  electrons propagating through the texture. The observation of the THE in the skyrmion phase of B20 compounds such as MnSi and FeGe has been pivotal in triggering the intensive research on skyrmionic systems \cite{Neubauer2009,Porter2014}. Remarkably, alongside its experimental observation, consistent theoretical description of the THE based on semiclassical arguments has been achieved recently based on {\it ab-initio} description of Mn$_{1-x}$Fe$_x$Si and Mn$_{1-x}$Fe$_x$Ge alloys \cite{Franz2014,Gayles2015,Freimuth2013}. 
Given that the locally antiferromagnetic order is more challenging to image experimentally, the analogue of the THE for AFM skyrmions would serve as an ideal tool for their detection. The formulation of such an effect for AFM skyrmions in analogy to the THE for ferromagnetic skyrmions is, however, by far more difficult, owing to more complicated dynamics of electrons in AFM textures even in the adiabatic limit, as well as different gauge symmetry of the Berry phase description of Bloch states in antiferromagnets \cite{Wilczek1984,Culcer2005,Shindou2005,Cheng2012,Gomonay2015}. 

In this work, based on the SU(2) semiclassical framework in combination with density functional theory (DFT) description of the electronic structure, we consider the adiabatic transport in AFM skyrmionic textures. Taking as an example synthetic AFMs, we predict that, while the transverse charge current is vanishing in AFM skyrmionic lattices subject to an external electric field, a strong response in terms of a transverse spin current could be used to detect the emergence of chiral lattices in AFM systems. 
We demonstrate the emergence of the corresponding topological {\it spin} Hall effect (TSHE) \cite{Yin2015} taking Fe/Cu/Fe trilayer as a representative example for a synthetic AFM system.
We discuss the physics of this new phenomenon, emphasize the importance of SU(2) gauge symmetry for its magnitude, and analyze its microscopics with respect to various parameters which determine the electronic structure of complex materials with the utter motivation of triggering the experimental observation of AFM skyrmions.

\

{\bf Conceptual and Computational Framework}. Within our method we access the transverse transport properties of two-dimensional synthetic AFMs with artificially imposed large-scale skyrmionic magnetic texture, based on the density-functional theory description of the electronic structure of a locally collinear system. The magnetic texture is considered to be given, constant in time, and determined by the spatially varying vector of the staggered magnetization $\boldsymbol{n}(\boldsymbol{r})$. Its influence on the dynamics of conduction electrons in real and reciprocal space is accounted for by considering effective semiclassical equations which govern the dynamics of wavepackets. For this purpose we use the formalism by Cheng and Niu \cite{Cheng2012} which takes into account a more complex SU(2) gauge freedom of doubly-degenerate electronic states in AFM crystals. 

For the case of doubly degenerate bands in locally collinear AFM crystals without spin-orbit interaction the degenerate Bloch states can be expressed as pure spinors $\ket{u_{a}}=\ket{A(\boldsymbol{k})}\ket{\uparrow(\boldsymbol{r})}$ and $\ket{u_{b}}=\ket{B(\boldsymbol{k})}\ket{\downarrow(\boldsymbol{r})}$, both of which are used to construct the electronic wavepacket. A key parameter in our theory is the overlap $\xi(\boldsymbol{k})=\braket{A|B}\in\left[0,1\right]$ describing the coupling between the $\boldsymbol{k}$ dependent parts of the spin-polarized Bloch states.
In contrast to conventional non-degenerate U(1) gauge symmetry the pivotal role that the parameter $\xi$ plays is the manifestation of the non-Abelian SU(2) gauge symmetry of the problem. Constructing the wavepacket's effective Lagrangian and using the variational principle, equations determining the adiabatic evolution of wavepacket's center in real ($\boldsymbol{r}$) and reciprocal ($\boldsymbol{k}$) spaces as well as its spin expectation value ($\boldsymbol{s}$) in the presence of a texture, can be derived \cite{Cheng2012,Xiao2010}. 
Without spin-orbit coupling and in the presence of an external electric field $\boldsymbol{E}$ these equations read (in atomic units):
\begin{subequations} 
	\label{eq:EOM}
	\begin{align}
	\label{eq:EOM_sdot}
	\dot{\boldsymbol{s}}&=(1-\xi^2)(\boldsymbol{s}\cdot\boldsymbol{n}) \dot{\boldsymbol{n}},\\
	\label{eq:EOM_kdot}
	\dot{\boldsymbol{k}}&=-(1-\xi^2)(\boldsymbol{s}\cdot\boldsymbol{n}) (\dot{\boldsymbol{r}}\times \boldsymbol{B}_{\mathrm{e}})-\boldsymbol{E} ,\\
	\label{eq:EOM_rdot}
	\dot{\boldsymbol{r}}&=\nabla_{\boldsymbol{k}}\epsilon-\frac{1}{2}((\boldsymbol{s}\times\boldsymbol{n})\cdot \dot{\boldsymbol{n}})\nabla_{\boldsymbol{k}}\ln(\xi).
	\end{align} 
\end{subequations}
According to Eq.~(\ref{eq:EOM_sdot}), in systems with SU(2) gauge group, the value of propagating electron's spin is not restricted to $\pm1$ in the local frame with the spin-quantization axis along $\boldsymbol{n}$, but it rather resides on a prolate spheroid with equatorial radius $\xi$ and its semi-major axis along $\boldsymbol{n}$ with length of $1$. This geometrical property, which allows for a misalignment between the spin and local direction of the magnetization in a texture, stands in sharp contrast to the usual treatment within the U(1) adiabatic dynamics. 
And while the alignment of $\boldsymbol{s}$ and $\boldsymbol{n}$ is restored for $\xi=0$ in the ``decoupled" limit of two effective U(1) antiparallel spin subsystems, in the other limit of $\xi\to1$ the propagating wavepacket's spin remains constant and independent of the spin texture. Another effect of $\xi$ lies in scaling the magnitude of the AFM analog of the ``emergent" magnetic field $\boldsymbol{B}_{\rm e,i}=\frac{1}{2}\varepsilon^{ijk}\boldsymbol{n}\cdot(\partial_j\boldsymbol{n}\times\partial_k\boldsymbol{n})$ in (\ref{eq:EOM_kdot}), which exerts a Lorentz force on propagating electrons \cite{Nagaosa2013,Cheng2012} and ultimately results in the TSHE. Additionally, in the last term of  Eq.~(\ref{eq:EOM_rdot}) the gradient of $\xi$ in $\boldsymbol{k}$-space modifies the group velocity. The latter effect originates in the mixed Berry curvature and is entirely absent in U(1)-systems \cite{Cheng2012}.

%========================================================
\begin{figure}[t!]
	\centering
	\includegraphics[angle=0,scale=0.92]{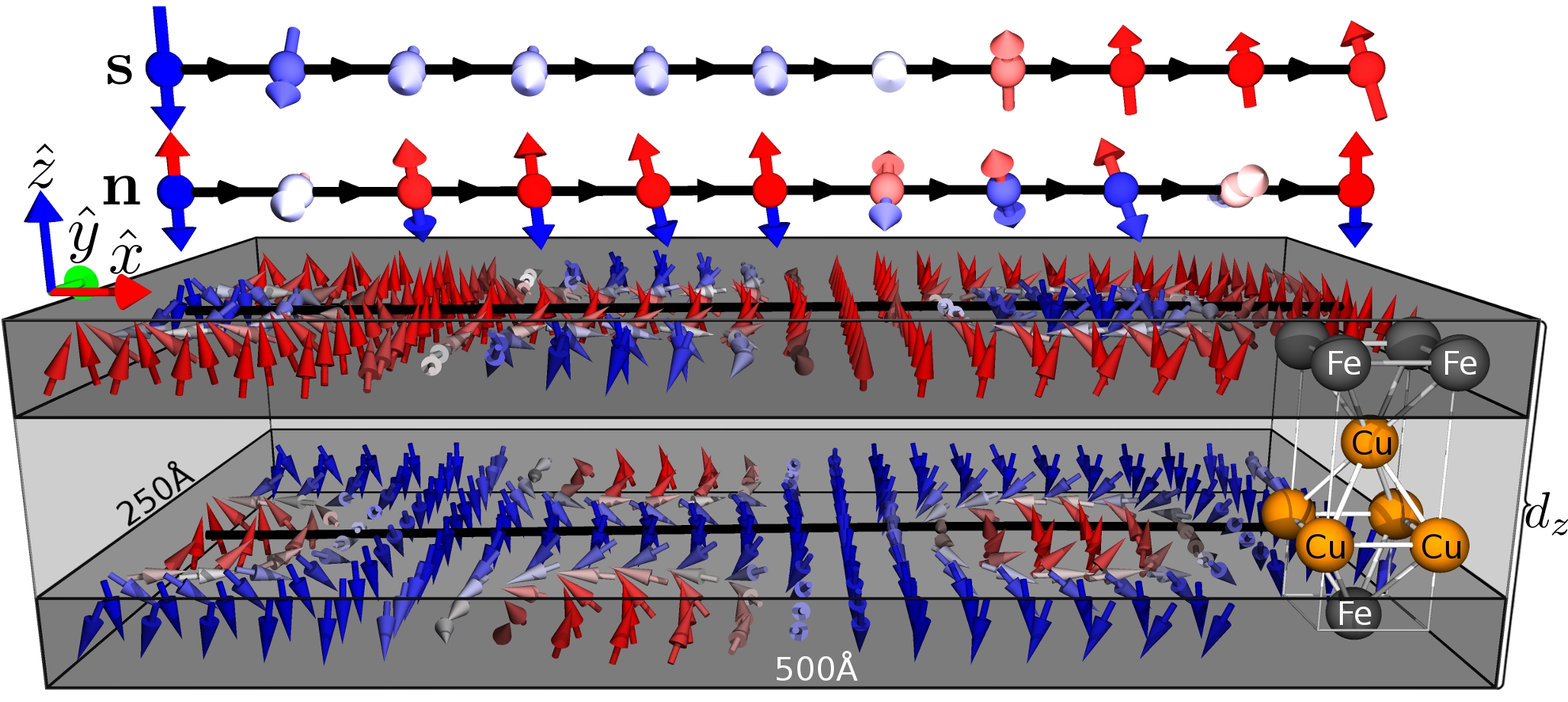}
	\caption{
		Magnetic texture and atomic structure of considered antiferromagnetic coupled trilayer \ce{Fe}/\ce{Cu2}/\ce{Fe}. The scales along the $z$ direction and in the plane are different. The evolution of spin $\boldsymbol{s}$ with respect to the local direction of staggered magnetization $\boldsymbol{n}$ is shown with arrows along the real-space trajectory (black lines) of a representative state at the Fermi surface with $\xi=0.6$ (for details see text).
	}
	\label{fig:schem}
\end{figure}
%========================================================

%========================================================
\begin{figure*}[t!]
	\centering
	\includegraphics[width=0.9\textwidth]{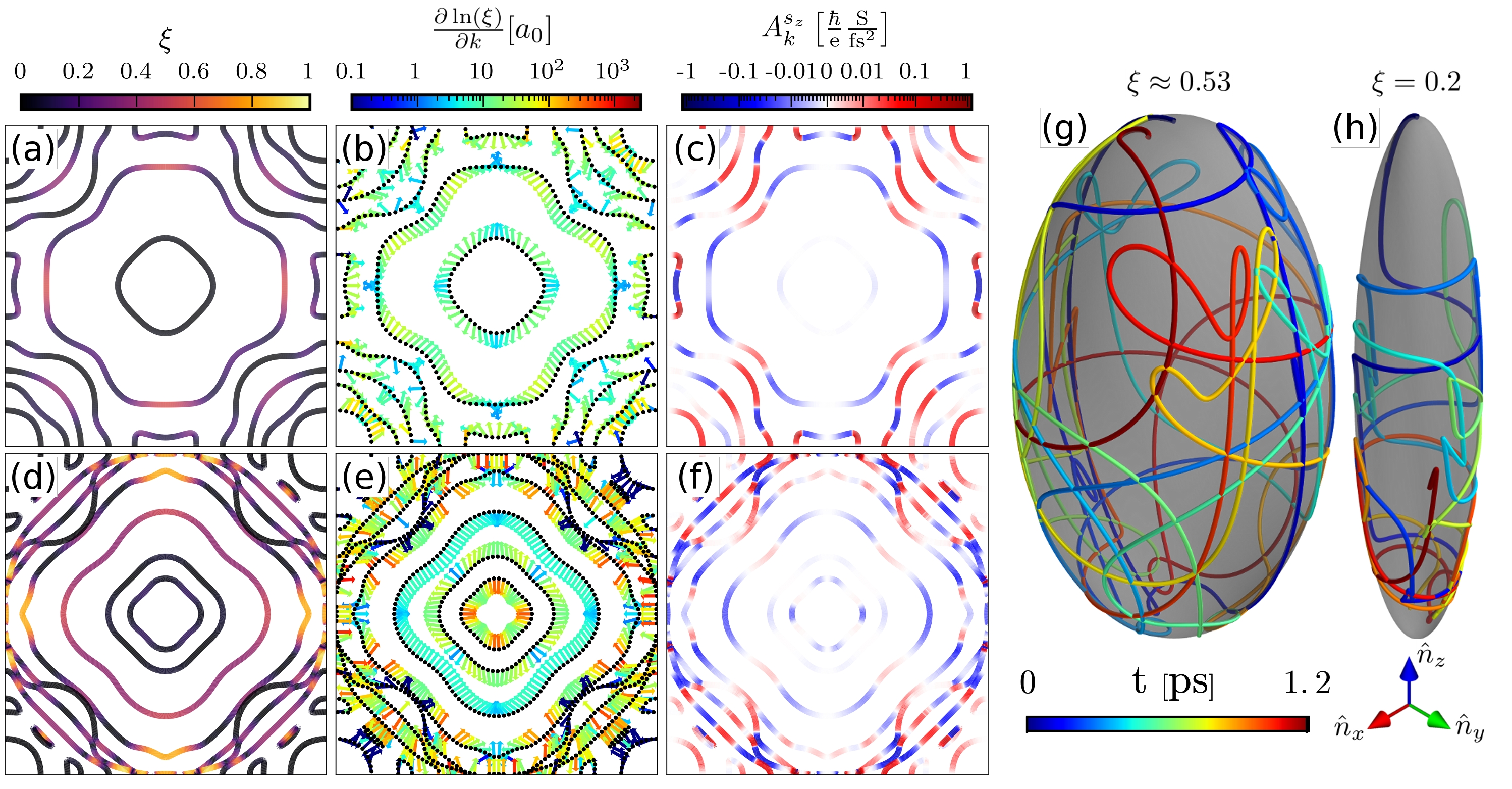}
	\caption{Left: Fermi surface resolved  $\xi$ (a,d), $\nabla_{\boldsymbol{k}}\ln(\xi)$ (b,e), and $A_{k}^{s_z}$ (c,f) for \ce{Fe}/\ce{Cu2}/\ce{Fe} (top row) and \ce{Fe}/\ce{Cu6}/\ce{Fe} (bottom row) trilayers. The direction of $\nabla_{\boldsymbol{k}}\ln(\xi)$ is displayed by arrows of unit length, while their color reflects the corresponding magnitude. $A_{k}^{s_z}$ is displayed on a symmetric logarithm scale, so that both positive and negative contributions are visible.
		Right: Evolution of spin of a representative state at the Fermi surface of \ce{Fe}/\ce{Cu2}/\ce{Fe}  in the frame of $\boldsymbol{n}$ plotted on a transparent prolate spheroid indicating the spin's codomain, for two values of the overlap $\xi$ of 0.2 (set manually) and 0.53 ({\it ab-initio} value).  The time-dependence is indicated with color.
	}
	\label{fig:xi}
\end{figure*}
%========================================================

Here, we are after the transport properties of an AFM skyrmion lattice in a thin film of a synthetic antiferromagnet, see Fig.\ref{fig:schem}. By referring to the linearized Boltzmann equation within the constant relaxation time approximation \cite{Franz2014} we utilize the equations for wavepacket propagation (\ref{eq:EOM}) to arrive at the expressions for the conductivities: 
\small
\begin{subequations} 
	\label{eq:cond}
	\begin{align}
	\label{eq:cond_1}
	\sigma_{ij}^{(1)}&=-\tau d_z \int d\boldsymbol{k} \frac{\partial f}{\partial \epsilon} \frac{\partial \epsilon}{\partial k_j} \left[\dot{r}_i\right],\\
	\label{eq:cond_2}
	\sigma_{ij}^{(2),\boldsymbol{s}}&=\tau^2 d_z \int d\boldsymbol{k} \frac{\partial f}{\partial \epsilon} (1-\xi^2) m_{jl} \left[\boldsymbol{s}_{\boldsymbol{n}}(\boldsymbol{s}\cdot\boldsymbol{n})(\dot{\boldsymbol{r}}\times\boldsymbol{B}_{\rm e})_l\dot{r}_i\right],
	\end{align} 
\end{subequations}
\normalsize
where $\sigma_{ij}^{(1)}$ and $\sigma_{ij}^{(2),\boldsymbol{s}}$ are the electrical and spin conductivity in first and second order with respect to the relaxation time $\tau$, respectively. The conductivities $\sigma_{ij}^{(1),\boldsymbol{s}}$ and $\sigma_{ij}^{(2)}$ are respectively obtained by including or excluding  from the corresponding expressions the spin in the frame of $\boldsymbol{n}$, $\boldsymbol{s}_{\boldsymbol{n}}$. 
We use the band dispersion $\epsilon(\boldsymbol{k})$ to compute the  effective mass tensor $m_{jl}=\frac{\partial^2 \epsilon}{\partial k_j\partial k_l}$.
The square brackets stand for the real-space averaging over the unit cell of the skyrmionic lattice and the implicit time dependence arising from the spin evolution. Parameter $d_z$ stands for the thickness of the antiferromagnetic thin film  (see Fig.\ref{fig:schem}). Equations (\ref{eq:cond}) allow us to study the transport properties of large scale AFM skyrmions in terms of the diagonal charge current, Eq.~(\ref{eq:cond_1}), and transverse spin Hall current originating in the emergent magnetic field due to the spin texture, Eq.~(\ref{eq:cond_2}).

According to (\ref{eq:cond}), the conductivities are intrinsically time-dependent and time-integrated quantities, since corresponding velocities and spin expectation values are found by evoluting the trajectories of the wavepackets in real space over time according to~(1). We evolute these equations for each of the states at the Fermi surface of the system, and for each of these $\boldsymbol{k}$-states  we initially set the real-space positions of the wavepackets on a dense mesh of points which cover the two-dimensional magnetic unit cell of the system, at the same time aligning the initial spin of the wavepacket either parallel or antiparallel to the local direction of $\boldsymbol{n}$. Within the formalism that we use we are ultimately
after the time-converged values of conductivities, although such effects as the sample size, the size of the chiral structure, as well as effects associated with impurity scattering
and wavepacket decoherence give physical meaning to the values of the conductivities at intermediate times.
At each of the time evolution steps for all systems studied we obtained vanishing values of transverse charge conductivities (i.e.~vanishing THE), in accord to the symmetry of the problem. The transverse spin signal (i.e.~TSHE) appears only in the second order with respect to $\tau$, and we thus refer to the corresponding conductivity as $\sigma_{ij}^{s_p}$.

Numerically, the Fermi surfaces, the band dispersion $\epsilon$, its derivatives, the values of parameter $\xi$ on the Fermi surface and its derivative $\nabla_{\boldsymbol{k}} \ln{\xi}$, were accurately evaluated for the collinear state of considered systems from {\it ab-initio} employing the technique of Wannier interpolation \cite{Mostofi2014}. Additionally, a model describing the skyrmion lattice texture has to be chosen in order to define the real-space distribution of $\boldsymbol{n}$ and $\boldsymbol{B}_{\rm{e}}$.
Provided this input, the equations for propagation of wavepackets~(1)~were evoluted to arrive at the values of charge and spin conductivities. This overall  procedure is motivated by the fact that for \ce{MnSi} theoretical predictions based on a semiclassical adiabatic U(1) formalism yield results which are in good agreement to experimental measurements \cite{Franz2014}. 

\

{\bf Topological Spin Hall Effect in Fe/Cu(001)/Fe trilayers}. In this work we focus on thin \ce{Fe}/\ce{Cu}(001)$_n$/\ce{Fe} trilayers,  with one \ce{Fe} layer on each side of the slab and varying \ce{Cu} thickness $n$, taking them as characteristic representatives of the class of synthetic AFMs where the AFM coupling is mediated by RKKY-type exchange \cite{Bennett1991}. For each thickness of the Cu spacer (varied between 0 and 6 layers), we assume the collinear AFM arrangement between the upper and lower Fe overlayers, keeping the coupling within each overlayer ferromagnetic, see Fig.\ref{fig:schem}. 
While we do not necessarily expect the emergence of AFM skyrmions in this particular family of systems, we consider them as exhibiting a typical electronic structure of systems for which the two-dimensional magnetic textures could be stabilized. 
The skyrmion lattice texture is imprinted by forcing the staggered magnetization $\boldsymbol{n}$ to vary in accordance to the so-called $3\boldsymbol{q}$-state with a wavelength of $190\,\textup{\AA}$, as observed for \ce{MnSi} \cite{Pfleiderer2009}, see Fig.~\ref{fig:schem}.
Since, to our knowledge, systems exhibiting AFM skyrmions have not been found to date, the setup we consider is designed with the purpose of exploring the impact of the SU(2) gauge freedom on transport properties in realistic systems in order to trigger an experimental observation of two-dimensional AFM textures. 
Within the choice of our axes, see Fig.\ref{fig:schem}, we thus study in the following the transverse spin $\sigma_{xy}^{s_z}$ and diagonal charge $\sigma_{xx}$ conductivities of the Fe/Cu(111)/Fe trilayers in the AFM skyrmion state. 

As follows from~Eq.~(\ref{eq:EOM}) and (\ref{eq:cond}) the values of overlap $\xi$ and its $\boldsymbol{k}$-derivative are of utter importance for the details of the wavepacket's evolution and resulting values of the conductivities. As apparent from the Fermi surface distributions, presented in Fig.\ref{fig:xi} for \ce{Fe}/\ce{Cu2}/\ce{Fe} and \ce{Fe}/\ce{Cu6}/\ce{Fe}, the behavior of $\xi$ and its gradient, which reflects the $\boldsymbol{k}$-dependent variation of the AFM hybridization between the states, can be very complex in thin films of transition-metals. The analysis of the thickness dependence of this behavior indicates that while the overlap $\xi$ between the $\ce{Fe}$-dominated states decreases with increasing \ce{Cu} thickness, the overall distribution of $\xi$ becomes more intricate. This effect can be attributed to the important role that the \ce{Cu} $s$-states play for the mediation of the AFM coupling between the Fe overlayers: while \ce{Cu}-states mostly carry moderate values of $\xi$, at the Brillouin zone boundary the \ce{Fe} and \ce{Cu} states hybridize strongly resulting in states with large $\xi$. Since $\xi$ vanishes exactly at the Brillouin zone boundary, very large values of $\nabla_{\boldsymbol{k}}\ln(\xi)$ are to be expected in its vicinity. 

To illustrate a profound influence that the finite value of $\xi$ has on the time-evolution of spin of a given wavepacket, in Fig.~\ref{fig:xi}(g,h) we show the spin evolution of the Fermi surface state at $\boldsymbol{k}\approx(0.74,0.07)\frac{\pi}{a}$ in the local frame of $\boldsymbol{n}$ as it propagates through the texture ($a$ is the in-plane lattice constant). We compare the case of computed from {\it ab-initio} value of $\xi\approx0.53$, with the case where we manually set $\xi$ to 0.2 (while keeping the small value of $|\nabla_{\boldsymbol{k}}\ln(\xi)|$ constant). 
As clearly visible in the figure, the wavepacket's spin displays a strong deviation from local $\boldsymbol{n}$ in both cases, even exhibiting a reversal as a function of time.  However, as intuitively expected from the equations of motion, the spin dynamics for a state with smaller value of $\xi$ is generally slower in the sense that the state's spin is more inclined to follow the texture's changes. 
The complex behavior of wavepacket's spin in response to non-zero overlap $\xi$ stands in sharp contract to the U(1) adiabatic picture valid for ferromagnetic skyrmions \cite{Cheng2012}.

Next, we analyse the Fermi surface distribution of the transverse spin conductivity at times approaching convergence.
The $\boldsymbol{k}$-dependence of $\sigma_{xy}^{s_z}$ can be characterized in a $\tau$-independent way by referring to the quantity $A_{k}^{s_z}$, defined through $\sigma_{xy}^{s_z}=\tau^2\int_{\rm{FS}} A_{k}^{s_z} \rm{d}k$. 
The Fermi surface distribution of $ A_{k}^{s_z}$ is shown for \ce{Fe}/\ce{Cu2}/\ce{Fe} and \ce{Fe}/\ce{Cu6}/\ce{Fe} in Fig.~\ref{fig:xi}(c,f). 
Their general feature is a pronounced compensation of positive and negative contributions which vary rapidly in magnitude. 
Each closed Fermi surface line comprises both positive and negative parts which makes the band-resolved analysis of  $\sigma_{xy}^{s_z}$ difficult. 
The effect of the SU(2) gauge freedom on the TSHE is best visible from comparing the values of $A_{k}^{s_z}$ for the coupled case of non-zero $\xi$ to the case when we put  $\xi$ to zero (not shown).
In the latter case  the equations of motion are effectively reduced to two copies of anti-aligned $U(1)$ systems.
For all considered systems the values of $A_{k}^{s_z}$ were consistently larger in the uncoupled case, while the difference in $A_{k}^{s_z}$ for both cases is closely correlated with the distribution of $\xi$ at the Fermi surface. 
This observation is in accord to an intuitive expectation that in the coupled case the effective ``dephasing" of the propagating electron's spin with respect to the direction of staggered magnetization is responsible for the suppression of the spin-polarization of transverse spin current, while this mechanism is absent in the decoupled 2$\times$U(1) case.

%========================================================
\begin{figure}[t!]
	\centering
	\includegraphics[angle=0,scale=0.92]{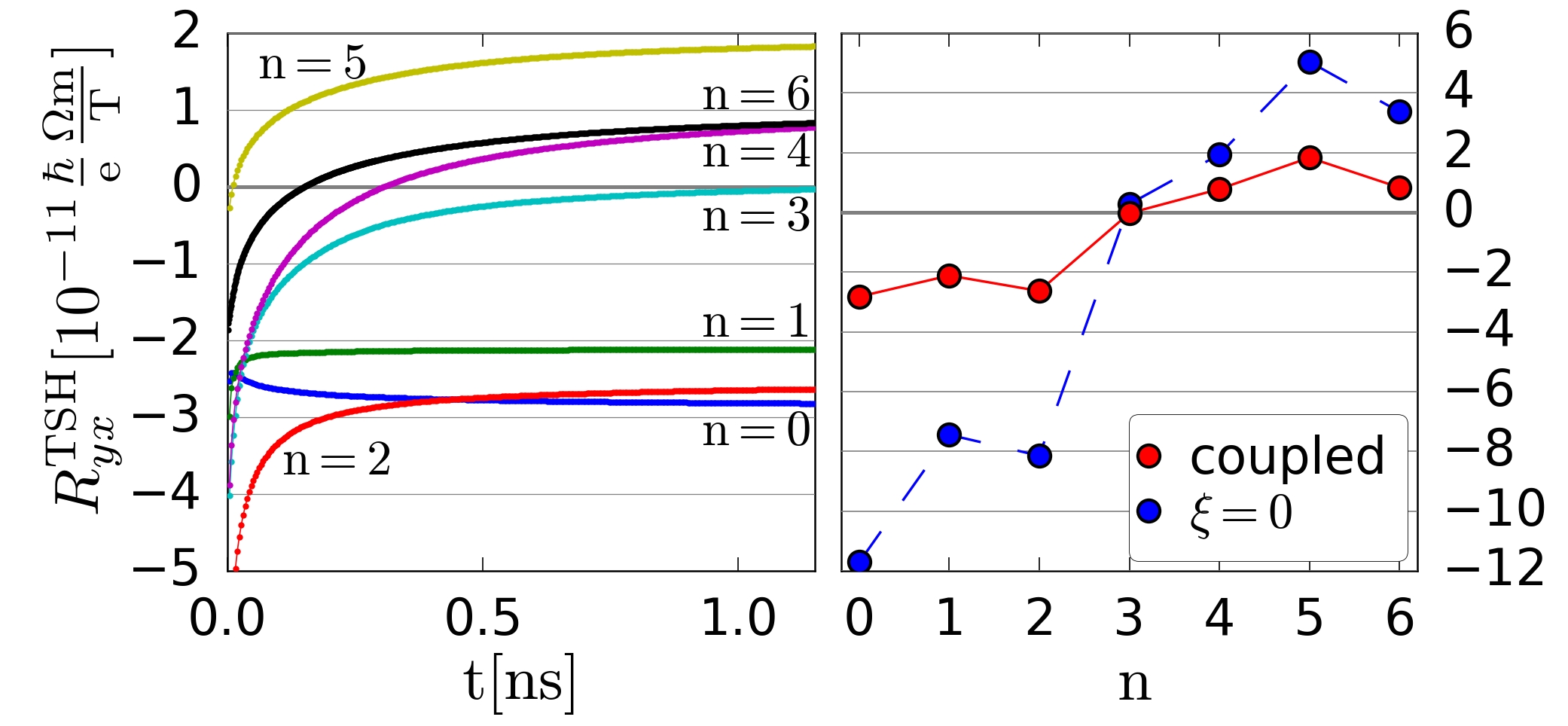}
	\caption{Left: Time-dependence of the TSH constant $R^{TSH}_{yx}$ in \ce{Fe}/\ce{Cu_n}/\ce{Fe} trilayers ($n$=0...6) after the electric field is applied to the system along the $x$-axis. Right: The dependence of the converged in time $R^{TSH}_{yx}$ as a function of Cu layer thickness for the coupled ($\xi\neq 0$) and uncoupled ($\xi=0$) cases.
	}
	\label{fig:timedep}
\end{figure}
%========================================================

In order to quantify the TSHE, in analogy to ferromagnetic skyrmions in Mn$_x$Fe$_{1-x}$Si and Mn$_x$Fe$_{1-x}$Ge alloys \cite{Franz2014},  we introduce the so-called topological spin Hall constant:
\begin{equation} \label{RTSH}
R^{\rm{TSH}}_{yx}=\frac{\sigma_{xy}^{s_z}}{\sigma_{xx}^{2} {\langle B_{\rm e}\rangle}},
\end{equation} 
where $\langle B_{\rm e}\rangle$ is defined as the averaged magnitude of the emergent field over the unit cell of the skyrmion lattice. The motivation behind introducing this constant lies in the observation that, while being a $\tau$-independent quantity, $R^{\rm{TSH}}_{yx}$ only weakly depends on the details of the real-space distribution of the emergent field. The TSH constant is thus a basic characteristic of an antiferromagnet which quantifies its topological spin Hall response as the texture with non-vanishing distribution of $\boldsymbol{B}_{\rm e}$ is imposed in some way. 
The TSH constant is a time-dependent quantity $-$  a property inherited from $\sigma_{xy}^{s_z}$ $-$ and in Fig.~\ref{fig:timedep} we present its time dependence for all systems that we considered. Owing to the intricate time evolution of spin of the wavepackets, as exemplified in Fig.\ref{fig:xi}(g,h), we observe that $R^{\rm{TSH}}_{yx}$ varies strongly in time and even exhibits a change of sign for some of the trilayers. 
The converged values of the TSH constant are reached on the scale of 1\,ns, which roughly corresponds to $0.5$ mm assuming a typical group velocity of $500$ km/s.  
And while this indicates that for the exact magnitude of the TSHE in AFM transition-metal films the details of spin scattering can be of importance, the overall range of values that $R^{\rm{TSH}}_{yx}$ displays over time according to our calculations, is encouraging. Taking into account that the value of the topological Hall constant computed with similar methods for ferromagnetic MnSi of the order of $3.0\times10^{-11}$ $\Omega$m/T is in agreement to experimentally estimated value, we conclude that the predicted magnitude of the TSHE in \ce{Fe}/\ce{Cu_n}/\ce{Fe} trilayers is sizable, and can be measured experimentally.

%========================================================
\begin{figure}[t!]
	\centering
	\includegraphics[angle=0,scale=0.92]{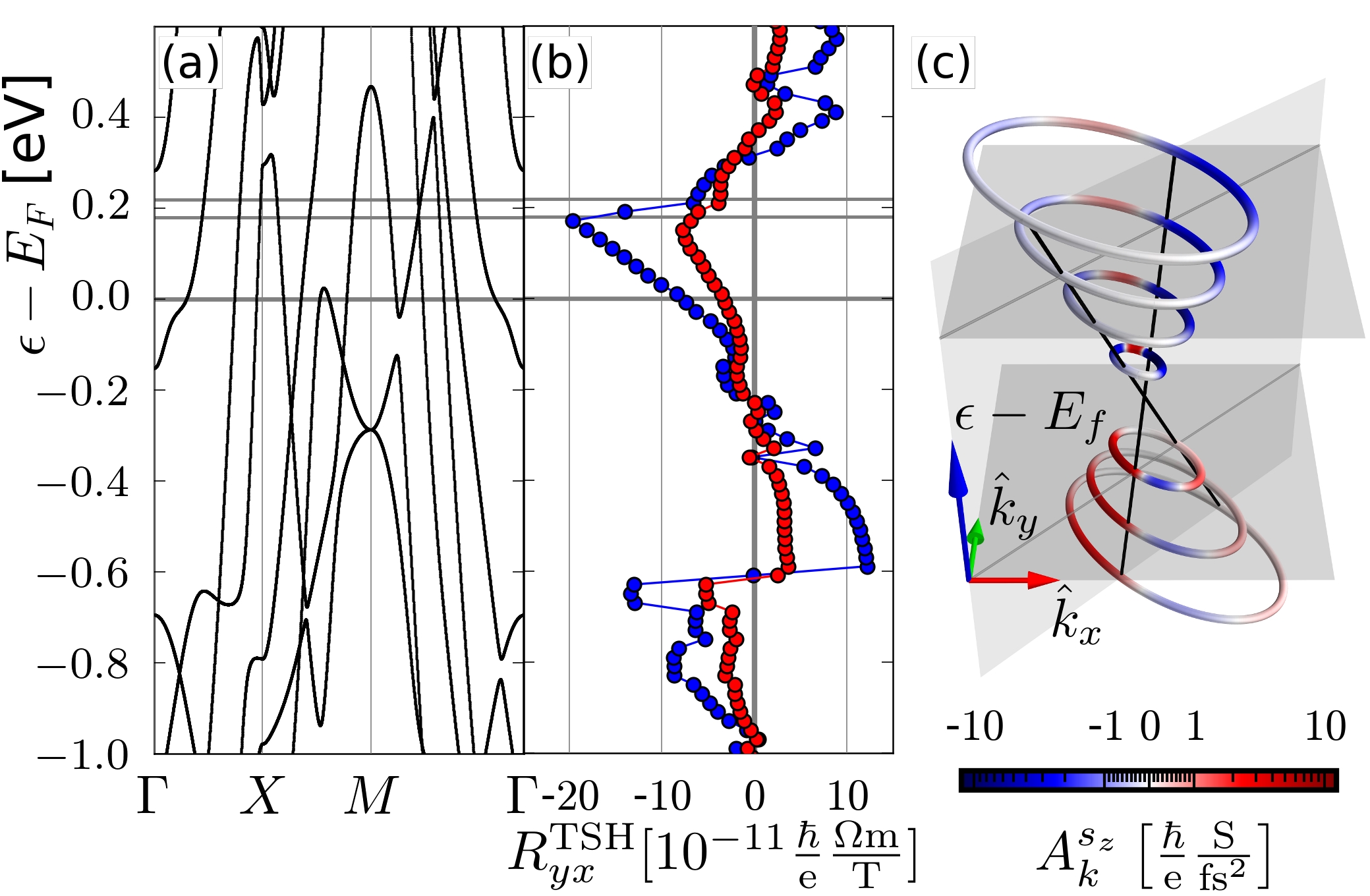}
	\caption{Properties of  \ce{Fe}/\ce{Cu2}/\ce{Fe}: (a) electronic band structure; (b)  energy dependence of $R^{\rm{TSH}}_{yx}$ for coupled ($\xi\neq 0$, red circles) and uncoupled ($\xi=0$, blue circles) cases; (c) the band contours of constant energy in the vicinity of a Dirac cone along $\Gamma $-$M$ line for energies from 0.16 to 0.23\,eV relative to the Fermi energy $E_F$. Gray horizontal planes at $0.18$ and $0.22$ eV correspond to horizontal thin lines in (a) and (b). Color on band contours shows the $\boldsymbol{k}$-resolved contributions to $\sigma^{s_z}_{xy}$ on a symmetric logarithmic scale. The Dirac bands along the high symmetry $\Gamma$-$M$ line are shown with thin black lines.
	}
	\label{fig:cond_energy_dep}
\end{figure}
%========================================================

The increasing trend of the TSH constant with the thickness of Cu spacer, presented in Fig.\ref{fig:timedep}, originates in the non-trivial evolution of the Fermi surface and the coupling between the states as $n$ is increased.
Additionally, in correspondence to the previously discussed behavior of the spin conductivity at the Fermi surface, the uncoupled case ($\xi=0$) yields noticeably larger values of the TSH constant. This does not necessarily have to be so, given the non-trivial oscillating in sign behavior of the $\boldsymbol{k}$-resolved spin conductivity over the Fermi surface contours. One can imagine a situation in which the large positive and negative $A_{k}^{s_z}$ on parts of the same Fermi surface line are modified in a different way as one goes from the coupled to the decoupled case, resulting in smaller $\xi=0$ value of the TSH constant (see for example the comparison of both cases for the value of the TSH constant at certain positions of the Fermi energy in \ce{Fe}/\ce{Cu_2}/\ce{Fe} in Fig.~\ref{fig:cond_energy_dep}(b)). 
Overall, taking into account that the separation between the Fe overlayers is pivotal for the magnitude of the exchange interaction which couples the upper texture to the lower one and thus has direct impact on the stability of such synthetic AFM skyrmions, the thickness of the spacer in synthetic AFMs provides a good handle of thermodynamics and transport properties of skyrmions in AFM systems.

Finally, we uncover the crucial role of the fine details of the electronic structure of transition-metal synthetic AFMs for the topological spin Hall effect. As visible in Fig.~\ref{fig:cond_energy_dep}(b), the value of the TSH constant is extremely sensitive to the position of the Fermi energy within the electronic structure of, for example, \ce{Fe}/\ce{Cu_2}/\ce{Fe}, Fig.~\ref{fig:cond_energy_dep}(a). As apparent from this plot, although $R^{\rm{TSH}}_{yx}$ is behaving very smoothly around the true Fermi energy of the system, abrupt step-wise changes can occur at special energies which correspond to the appearance or disappearance of the band edges. Another reason for a large variation of the TSH constant is passing of the Fermi energy through Dirac points of band crossings. This second mechanism manifests prominently for example at the Dirac cone appearing along the $\Gamma-M$ direction at about $0.19\,\textup{eV}$ in energy, at which point both coupled and uncoupled $R^{\rm{TSH}}_{yx}$ change significantly.  By analyzing  the distribution of $A_{k}^{s_z}$ near this Dirac crossing, Fig.~\ref{fig:cond_energy_dep}(c), we can state that it is the rapid increase in $A_{k}^{s_z}$ as the Fermi energy approaches the Dirac point combined with the change in its sign as the Dirac point is crossed in energy, which makes such a drastic influence on the value of the TSH constant.
This behavior is rather reminiscent of that of the Berry curvature of the electronic states in $\boldsymbol{k}$-space, also exhibiting rapid variation as the Fermi energy is passing through the points of degeneracy and leading to a strong energy dependence of the anomalous Hall effect \cite{Fang2003}.   
At the end, we conclude that the TSHE is extremely susceptible to the changes in the electronic structure of materials, which thus makes it an ideal property for electronic structure engineering.

\

{\bf Conclusions}. In this work, employing the semiclassical framework grounded in the {\it ab-initio} description of the 
electronic structure we have demonstrated the emergence of sizeable transverse spin currents in candidate AFM skyrmionic systems. 
Our findings imply that the spin currents of such magnitude can be detected experimentally, resulting thus in an experimental observation of the AFM counterpart of the THE $-$ the topological spin Hall effect. 
The discovery of the TSHE in AFM skyrmions which we discuss here opens new vistas for the detection of AFM skyrmions in various systems, and we believe that the TSHE will play the role similar to that of the conventional THE for the discovery of ferromagnetic skyrmions. 
Moreover, the prominent TSHE marks the AFM skyrmions as promising objects for pure spin current generation, which, as opposed to the spin Hall effect in paramagnets, is not relying on relativistic effects in the electronic structure. We suggest that such spin currents can be detected with standard techniques, i.e. by detecting the spin-torque that topological spin Hall currents would exert on a ferromagnet brought in contact with the AFM skyrmionic system, or by magneto-optical means in terms of corresponding spin accumulation at the edges of the sample. 
Concerning the prospect for theoretical developments of the approach that we use, the most natural conceptual extension would lie in including the effect of momentum and spin scattering on the details of formation and propagation of transverse spin currents in complex chiral antiferromagnetic media with disorder, which would allow reliable estimation of the TSHE in realistic disordered materials with the aim of scrutinizing  the influence of finite temperature and impurity scattering on the properties of such ``chiral" spin currents.

\

{\bf Methods}. The {\it ab-initio} calculations were performed in the generalized gradient approximation to the exchange correlation functional using the film version of the full-potential linearized method as implemented in the J\"ulich code {\tt FLEUR} \cite{fleur}.
The $z$-positions of the atoms were relaxed starting from the in-plane lattice parameter $a\approx4.59 a_0$, where $a_0$ is Bohr's radius. Subsequently the charge  densities were converged with at least 300 $\boldsymbol{k}$-points. 
Maximally localized Wannier functions from \ce{Fe} and \ce{Cu} $s,p,d$ initial states were constructed on a $12\times12$ grid using the wannier90 code \cite{Mostofi2014}. A frozen window maximum was chosen around 2.3\,eV above the Fermi energy. The systems' Hamiltonian, spin, and overlap $\xi$ were obtained in the space of maximally localized Wannier functions via an interface between FLEUR and wannier90. 
For the transport calculations we employed the classical Runge-Kutta method enforcing a maximal real space step of $10 a_0$ while generally using a set time step $\Delta t\approx 1\textup{fs}$. 
An initial $10\times10$ real space grid spanning the magnetic unit cell was used, larger grids did not yield a significant difference.
In the Fermi energy calculations (Fig. \ref{fig:timedep}) we started from $16386$ $\boldsymbol{k}$-points, for the energy dependence (Fig. \ref{fig:cond_energy_dep}) we used $4096$.

\

{\bf Acknowledgments}. We would like to thank Stefan Heinze and Helen Gomonay for numerous discussions on the physics of antiferromagnetic skyrmions. This work has been supported by the Deutsche Forschungsgemeinschaft (DFG) through the Collaborative Research Center SFB 1238.
We acknowledge funding from the European Union's Horizon 2020 research and innovation programme under grant agreement number 665095 (FET-Open project MAGicSky).
We especially acknowledge the computing time on the supercomputers JUQUEEN and JURECA at J\"ulich Supercomputing Centre and JARA-HPC of RWTH Aachen University.

\providecommand{\WileyBibTextsc}{}
\let\textsc\WileyBibTextsc
\providecommand{\othercit}{}
\providecommand{\jr}[1]{#1}
\providecommand{\etal}{~et~al.}


\begin{thebibliography}{[10]}

\bibitem{Pfleiderer2009}% article
 \textsc{S.~M{\"{u}}hlbauer},  \textsc{B.~Binz},  \textsc{F.~Jonietz},
  \textsc{C.~Pfleiderer},  \textsc{A.~Rosch},  \textsc{A.~Neubauer},
  \textsc{R.~Georgii},  and  \textsc{P.~B{\"{o}}ni}\iffalse {Skyrmion lattice
  in a chiral magnet.}\fi,
 \jr{Science} \textbf{323}(5916), 915--919 (2009).


\bibitem{Yu2011}% article
 \textsc{X.\,Z. Yu},  \textsc{N.~Kanazawa},  \textsc{Y.~Onose},
  \textsc{K.~Kimoto},  \textsc{W.\,Z. Zhang},  \textsc{S.~Ishiwata},
  \textsc{Y.~Matsui},  and  \textsc{Y.~Tokura}\iffalse {Near room-temperature
  formation of a skyrmion crystal in thin-films of the helimagnet FeGe.}\fi,
 \jr{Nat. Mater.} \textbf{10}, 106--109 (2011).


\bibitem{Moreau-Luchaire2016}% article
 \textsc{C.~Moreau-Luchaire},  \textsc{C.~Moutaﬁs},  \textsc{N.~Reyren},
  \textsc{J.~Sampaio},  \textsc{C.\,A.\,F. Vaz},  \textsc{N.~{Van Horne}},
  \textsc{K.~Bouzehouane},  \textsc{K.~Garcia},  \textsc{C.~Deranlot},
  \textsc{P.~Warnicke},  \textsc{P.~Wohlh{\"{u}}ter},  \textsc{J.\,M. George},
  \textsc{M.~Weigand},  \textsc{J.~Raabe},  \textsc{V.~Cros},  and
  \textsc{A.~Fert}\iffalse {Additive interfacial chiral interaction in
  multilayers for stabilization of small individual skyrmions at room
  temperature}\fi,
 \jr{Nat. Nanotech.} \textbf{11}(5), 444--448 (2016).


\bibitem{Soumyanarayanan2016}% article
 \textsc{A.~Soumyanarayanan},  \textsc{M.~Raju},  \textsc{A.\,L.\,G. Oyarce},
  \textsc{A.\,K.\,C. Tan},  \textsc{M.\,Y. Im},  \textsc{A.\,P. Petrovic},
  \textsc{P.~Ho},  \textsc{K.\,H. Khoo},  \textsc{M.~Tran},  \textsc{C.\,K.
  Gan},  \textsc{F.~Ernult},  and  \textsc{C.~Panagopoulos}\iffalse {Tunable
  Room Temperature Magnetic Skyrmions in Ir/Fe/Co/Pt Multilayers}\fi,
 \jr{arXiv}:1606.06034 (2016).


\bibitem{Woo2016}% article
 \textsc{S.~Woo},  \textsc{K.~Litzius},  \textsc{B.~Kr{\"{u}}ger},
  \textsc{M.\,Y. Im},  \textsc{L.~Caretta},  \textsc{K.~Richter},
  \textsc{M.~Mann},  \textsc{A.~Krone},  \textsc{R.\,M. Reeve},
  \textsc{M.~Weigand},  \textsc{P.~Agrawal},  \textsc{I.~Lemesh},
  \textsc{M.\,A. Mawass},  \textsc{P.~Fischer},  \textsc{M.~Kl{\"{a}}ui},  and
  \textsc{G.\,S.\,D. Beach}\iffalse {Observation of room-temperature magnetic
  skyrmions and their current-driven dynamics in ultrathin metallic
  ferromagnets}\fi,
 \jr{Nat. Mater.} \textbf{15}(5), 501--506 (2016).


\bibitem{Dupe2016}% article
 \textsc{B.~Dup{\'{e}}},  \textsc{G.~Bihlmayer},  \textsc{M.~B{\"{o}}ttcher},
  \textsc{S.~Bl{\"{u}}gel},  and  \textsc{S.~Heinze}\iffalse {Engineering
  skyrmions in transition-metal multilayers for spintronics}\fi,
 \jr{Nat. Commun.} \textbf{7}, 11779 (2016).


\bibitem{Purnama2015}% article
 \textsc{I.~Purnama},  \textsc{W.\,L. Gan},  \textsc{D.\,W. Wong},  and
  \textsc{W.\,S. Lew}\iffalse {Guided current-induced skyrmion motion in 1D
  potential well.}\fi,
 \jr{Sci. Rep.} \textbf{5}, 10620 (2015).


\bibitem{Zhang2016}% article
 \textsc{X.~Zhang},  \textsc{Y.~Zhou},  and  \textsc{M.~Ezawa}\iffalse
  {Antiferromagnetic Skyrmion: Stability, Creation and Manipulation}\fi,
 \jr{Sci. Rep.} \textbf{6}, 24795 (2016).


\bibitem{Zhang2016a}% article
 \textsc{X.~Zhang},  \textsc{Y.~Zhou},  and  \textsc{M.~Ezawa}\iffalse
  {Magnetic bilayer-skyrmions without skyrmion Hall effect}\fi,
 \jr{Nat. Commun.} \textbf{7}, 10293 (2016).


\bibitem{Velkov2016}% article
 \textsc{H.~Velkov},  \textsc{O.~Gomonay},  \textsc{M.~Beens},
  \textsc{G.~Schwiete},  \textsc{A.~Brataas},  \textsc{J.~Sinova},  and
  \textsc{R.\,A. Duine}\iffalse {Phenomenology of current-induced skyrmion
  motion in antiferromagnets}\fi,
 \jr{New J. Phys.} \textbf{18}(7), 075016 (2016).


\bibitem{Wadley2016}% article
 \textsc{P.~Wadley},  \textsc{B.~Howells},  \textsc{J.~Elezny},
  \textsc{C.~Andrews},  \textsc{V.~Hills},  \textsc{R.\,P. Campion},
  \textsc{V.~Novak},  \textsc{K.~Olejnik},  \textsc{F.~Maccherozzi},
  \textsc{S.\,S. Dhesi},  \textsc{S.\,Y. Martin},  \textsc{T.~Wagner},
  \textsc{J.~Wunderlich},  \textsc{F.~Freimuth},  \textsc{Y.~Mokrousov},
  \textsc{J.~Kune},  \textsc{J.\,S. Chauhan},  \textsc{M.\,J. Grzybowski},
  \textsc{A.\,W. Rushforth},  \textsc{K.\,W. Edmonds},  \textsc{B.\,L.
  Gallagher},  and  \textsc{T.~Jungwirth}\iffalse {Electrical switching of an
  antiferromagnet}\fi,
 \jr{Science} \textbf{351}(6273), 587--590 (2016).


\bibitem{Jungwirth2016}% article
 \textsc{T.~Jungwirth},  \textsc{X.~Marti},  \textsc{P.~Wadley},  and
  \textsc{J.~Wunderlich}\iffalse {Antiferromagnetic spintronics}\fi,
 \jr{Nat. Nanotech.} \textbf{11}(3), 231--241 (2016).


\bibitem{Wiesendanger2009}% article
 \textsc{R.~Wiesendanger}\iffalse {Spin mapping at the nanoscale and atomic
  scale}\fi,
 \jr{Rev. Mod. Phys.} \textbf{81}(4), 1495--1550 (2009).


\bibitem{Bruno2004}% article
 \textsc{P.~Bruno},  \textsc{V.\,K. Dugaev},  and
  \textsc{M.~Taillefumier}\iffalse {Topological Hall Effect and Berry Phase in
  Magnetic Nanostructures}\fi,
 \jr{Phys. Rev. Lett.} \textbf{93}(9), 096806 (2004).


\bibitem{Nagaosa2013}% article
 \textsc{N.~Nagaosa} and  \textsc{Y.~Tokura}\iffalse {Topological properties
  and dynamics of magnetic skyrmions.}\fi,
 \jr{Nat. Nanotech.} \textbf{8}(12), 899--911 (2013).


\bibitem{Denisov2016}% article
 \textsc{K.\,S. Denisov},  \textsc{I.\,V. Rozhansky},  \textsc{N.\,S.
  Averkiev},  and  \textsc{E.~L{\"{a}}hderanta}\iffalse {Electron Scattering on
  a Magnetic Skyrmion in the Nonadiabatic Approximation}\fi,
 \jr{Phys. Rev. Lett.} \textbf{117}(2), 027202 (2016).


\bibitem{Neubauer2009}% article
 \textsc{A.~Neubauer},  \textsc{C.~Pfleiderer},  \textsc{B.~Binz},
  \textsc{A.~Rosch},  \textsc{R.~Ritz},  \textsc{P.\,G. Niklowitz},  and
  \textsc{P.~B{\"{o}}ni}\iffalse {Topological hall effect in the a phase of
  MnSi}\fi,
 \jr{Phys. Rev. Lett.} \textbf{102}(18), 186602 (2009).


\bibitem{Porter2014}% article
 \textsc{N.\,A. Porter},  \textsc{J.\,C. Gartside},  and  \textsc{C.\,H.
  Marrows}\iffalse {Scattering mechanisms in textured FeGe thin films:
  Magnetoresistance and the anomalous Hall effect}\fi,
 \jr{Phys. Rev. B} \textbf{90}(2), 024403 (2014).


\bibitem{Franz2014}% article
 \textsc{C.~Franz},  \textsc{F.~Freimuth},  \textsc{A.~Bauer},
  \textsc{R.~Ritz},  \textsc{C.~Schnarr},  \textsc{C.~Duvinage},
  \textsc{T.~Adams},  \textsc{S.~Bl{\"{u}}gel},  \textsc{A.~Rosch},
  \textsc{Y.~Mokrousov},  and  \textsc{C.~Pfleiderer}\iffalse {Real-space and
  reciprocal-space Berry phases in the Hall effect of Mn(1-x)Fe(x)Si.}\fi,
 \jr{Phys. Rev. Lett.} \textbf{112}(18), 186601 (2014).


\bibitem{Gayles2015}% article
 \textsc{J.~Gayles},  \textsc{F.~Freimuth},  \textsc{T.~Schena},
  \textsc{G.~Lani},  \textsc{P.~Mavropoulos},  \textsc{R.\,A. Duine},
  \textsc{S.~Bl{\"{u}}gel},  \textsc{J.~Sinova},  and
  \textsc{Y.~Mokrousov}\iffalse {Dzyaloshinskii-Moriya Interaction and Hall
  Effects in the Skyrmion Phase of Mn1-xFexGe}\fi,
 \jr{Phys. Rev. Lett.} \textbf{115}(3), 036602 (2015).


\bibitem{Freimuth2013}% article
 \textsc{F.~Freimuth},  \textsc{R.~Bamler},  \textsc{Y.~Mokrousov},  and
  \textsc{A.~Rosch}\iffalse {Phase-space Berry phases in chiral magnets:
  Dzyaloshinskii-Moriya interaction and the charge of skyrmions}\fi,
 \jr{Phys. Rev. B} \textbf{88}(21), 214409 (2013).


\bibitem{Wilczek1984}% article
 \textsc{F.~Wilczek} and  \textsc{A.~Zee}\iffalse {Appearance of gauge
  structure in simple dynamical systems}\fi,
 \jr{Phys. Rev. Lett.} \textbf{52}(24), 2111--2114 (1984).


\bibitem{Culcer2005}% article
 \textsc{D.~Culcer},  \textsc{Y.~Yao},  and  \textsc{Q.~Niu}\iffalse {Coherent
  wave-packet evolution in coupled bands}\fi,
 \jr{Phys. Rev. B} \textbf{72}(8), 085110 (2005).


\bibitem{Shindou2005}% article
 \textsc{R.~Shindou} and  \textsc{K.\,I. Imura}\iffalse {Noncommutative
  geometry and nonabelian Berry phase in the wave-packet dynamics of Bloch
  electrons}\fi,
 \jr{Nucl. Phys. B} \textbf{720}(3), 399--435 (2004).


\bibitem{Cheng2012}% article
 \textsc{R.~Cheng} and  \textsc{Q.~Niu}\iffalse {Electron dynamics in slowly
  varying antiferromagnetic texture}\fi,
 \jr{Phys. Rev. B} \textbf{86}(24), 245118 (2012).


\bibitem{Gomonay2015}% article
 \textsc{O.~Gomonay}\iffalse {Berry-phase effects and electronic dynamics in a
  noncollinear antiferromagnetic texture}\fi,
 \jr{Phys. Rev. B} \textbf{91}(14), 144421 (2015).


\bibitem{Yin2015}% article
 \textsc{G.~Yin},  \textsc{Y.~Liu},  \textsc{Y.~Barlas},  \textsc{J.~Zang},
  and  \textsc{R.\,K. Lake}\iffalse {Topological spin Hall effect resulting
  from magnetic skyrmions}\fi,
 \jr{Phys. Rev. B} \textbf{92}(2), 024411 (2015).


\bibitem{Xiao2010}% article
 \textsc{D.~Xiao},  \textsc{M.\,C. Chang},  and  \textsc{Q.~Niu}\iffalse {Berry
  phase effects on electronic properties}\fi,
 \jr{Rev. Mod. Phys.} \textbf{82}(3), 1959--2007 (2010).


\bibitem{Mostofi2014}% article
 \textsc{A.\,A. Mostofi},  \textsc{J.\,R. Yates},  \textsc{G.~Pizzi},
  \textsc{Y.\,S. Lee},  \textsc{I.~Souza},  \textsc{D.~Vanderbilt},  and
  \textsc{N.~Marzari}\iffalse {An updated version of wannier90: A tool for
  obtaining maximally-localised Wannier functions}\fi,
 \jr{Comput. Phys. Commun.} \textbf{185}(8), 2309--2310 (2014).


\bibitem{Bennett1991}% article
 \textsc{W.\,R. Bennett},  \textsc{W.~Schwarzacher},  and  \textsc{W.\,F.
  Egelhoff}\iffalse {Concurrent enhancement of Kerr rotation and
  antiferromagnetic coupling in epitaxial Fe/Cu/Fe structures (abstract)}\fi,
 \jr{J. Appl. Phys.} \textbf{70}(10), 5881 (1991).


\bibitem{Fang2003}% article
 \textsc{Z.~Fang}\iffalse {The Anomalous Hall Effect and Magnetic Monopoles in
  Momentum Space}\fi,
 \jr{Science} \textbf{302}(5642), 92--95 (2003).


\othercit
\bibitem{fleur}% misc
http://www.flapw.de.


\end{thebibliography}
\end{document}